\documentclass[pra,twocolumn,superscriptaddress,showpacs]{revtex4}
\usepackage{graphicx}
\usepackage{amsmath}
\usepackage{amssymb,bm}
\usepackage{bm,hyperref}
\usepackage{CJK,color}

\begin{document}
\begin{CJK*}{UTF8}{gbsn}
\title{Interaction Effects on Wannier Functions for Bosons in Optical Lattice}
\author{Shaoqi Zhu(朱少奇)}
\affiliation{International Center for Quantum Materials, Peking University, 100871, Beijing, China}
\author{Biao Wu(吴飙)}
\affiliation{International Center for Quantum Materials, Peking University, 100871, Beijing, China}
\affiliation{Collaborative Innovation Center of Quantum Matter, Beijing 100871, China}
\date{\today}

\date{\today}

\begin{abstract}
We have numerically calculated the single band Wannier functions for interacting Bose gases in optical lattices with a self-consistent approach. We find that the Wannier function is broadened by repulsive atom interaction. The tunneling parameter $J$ and on-site interaction $U$ computed with the broadened Wannier functions are found to change significantly for different atomic number per site. Our theory can explain the nonuniform  atomic clock shift observed in [Campbell {\it et al.}, Science {\bf 313}, 649 (2006)].
\end{abstract}
\maketitle
\end{CJK*}

\section{Introduction}
In condensed matter physics,  people are often interested in the ground state of the system
and its low energy excitations. This allows people to focus only on the lowest bands of
the systems by mapping the system to a lattice model with 
Wannier functions~\cite{GW PR}. The Hubbard model is arguably the most famous
of all such lattice models~\cite{NA SSP}.  In addition,  Wannier functions 
are a more natural and better choice than Bloch waves for narrow-band materials in 
computational physics~\cite{NM PRB}. There is a lot of freedom to choose a set of
Wannier functions as basis and there has been great efforts to find the best Wannier 
functions~\cite{WK PRB}.  
However, all the discussion is done in the context of single-particle physics. 
The broadening of Wannier functions by repulsive interaction seems have never been
discussed in traditional condensed matter physics.

The situation starts to change with the development of ultracold atomic physics, in 
particular, with the experimental observation of superfluid to Mott transition in optical 
lattice~\cite{IB Nature,IB RMP}.  In the early theoretical treatment of such a system, 
the mapping from the realistic continuous  system to the Bose-Hubbard model is done
with single particle Wannier function~\cite{DJ PRL}. However, due to the simplicity of 
the periodic potential and small energy scales in the system, it is tempting to
think that the broadening of Wannier function by the on-site repulsive interaction 
may have significant effect on the system. There have been many theoretical 
efforts~\cite{JL NJP, ZL arXiv, PS PRA, DM PRA, VY LP, DL NJP} trying to 
give a good description of interaction effects on Wannier function. 

There are also strong experimental evidences on the the broadening of Wannier functions. 
In atomic clock experiment~\cite{KG PRL, GC Science, GC2 Science, MM Science}, atomic interaction is the main reason for frequency shift. In the experiment of $^{87}Rb$ atomic clock in optical lattice~\cite{GC Science}, nonuniform frequency shift was observed for different occupation number per site. This is clearly due to the broadening of Wannier function by the repulsive atomic interaction. 

In Ref. \cite{BW CPL}, a self-consistent approach is developed to take account 
of the interaction effect on Wannier function. The use of a different set of Wannier functions
will result in a different tunneling parameter $J$ and the on-site interaction $U$ for the lattice
model and thus a different ground state. The self-consistent approach in Ref. \cite{BW CPL}
uses  a general variational principle to choose the set of Wannier functions that minimizes the ground state energy of the lattice model. This method is in spirit the same as the 
MCTDHB theory~\cite{AS PRA, OA PRA, KS NJP}. 

In this work we use the self-consistent approach in Ref. \cite{BW CPL} to compute
the interaction broadened Wannier functions for a Bose gas in an optical lattice. 
We focus on both the superfluid regime and the Mott insulator regime. The Wannier functions
are used to calculate the tunneling parameter $J$ and the on-site interaction $U$ in 
the Bose-Hubbard model. They are found to be significantly affected by the $s$-wave
scattering length, lattice strength, and most importantly the number of particles per site. 
In the end,  we apply the approach to the experiment in Ref. ~\cite{GC Science}; our theoretical 
results match very well the experimental data. 

Our paper is organized as follows.  In section \uppercase\expandafter{\romannumeral2}, a quick review of the self-consistent approach is given in Ref. \cite{BW CPL}. In section \uppercase\expandafter{\romannumeral3} the nonlinear equations are solved in the superfluid regime for one dimensional optical lattice; the tunneling parameter $J$ and on-site interaction $U$ are calculated with the changing of lattice depth and interaction strength. In section \uppercase\expandafter{\romannumeral4}, we solve the nonlinear equations for the Mott insulator regime where the expansion basis are single particle Wannier functions; $J$ and $U$ are calculated accordingly. In section \uppercase\expandafter{\romannumeral5}, the theory is applied to the  experiment 
in Ref. ~\cite{GC Science} and a good agreement is found between theory and experiment.

\section{Self-Consistent Approach for Wannier functions}
In this section we give a brief summary on the self-consistent approach 
to compute the interaction effects on Wannier functions developed 
in Ref.\cite{BW CPL}.  We consider a  Bose gas where the weak atomic 
interaction can be well described by the $s$-wave scattering. The second quantized 
Hamiltonian for this kind of system  is given by 
\begin{equation}
\begin{aligned}
\hat{H}=&\int{d\bm{r}\hat{\psi}^{\dagger}(\bm{r})[-\frac{\hbar^2}{2m}\nabla^2+V(\bm{r})]\hat{\psi}(\bm{r})}\\
+&\frac{g_0}{2}\int{d\bm{r}[\hat{\psi}^{\dagger}(\bm{r})\hat{\psi}^{\dagger}(\bm{r})\psi(\bm{r})\psi(\bm{r})]}\,,
\end{aligned}
\end{equation}
where $m$ is the atomic mass, $V(\bm{r})$ is the external potential. We choose $V(\bm{r})=V_0sin(\pi \bm{r})^2$ in this context, where $V_0$ is the optical lattice strength. And $g_0=4\pi\hbar^2a_s/m$ is the interaction 
strength related to the $s$-wave scattering length $a_s$. The single-band approximation 
is to expand the bosonic field operator $\psi(\bm{r})$  as
\begin{equation}
\hat{\psi}(\bm{r})=\sum_j\hat{a}_jW_j(\bm{r}),
\end{equation}
where $W_j(\bm{r})=W(\bm{r}-\bm{r}_j)$ is the first band Wannier function at site $j$ and $\hat{a}_j$ is the associated annihilation operator. The ground state $|G_t\rangle$ in the single-band approximation can 
be generally written as $|G_t\rangle=F({\hat{a}_j}^\dagger)|{\rm vaccum}\rangle$, where $F$ 
is the function to be found by solving the resulted lattice model. The ground state energy
$E_G=\langle G_t|\hat{H}|G_t\rangle$ certainly changes with the choice of Wannier function
$W_j(\bm{r})$.  The best Wannier function is the one that  minimize the  
single-band ground state energy $E_G$. Mathematically, this is to do the following variation
\begin{equation}
\frac{\delta E_G}{\delta W^*(\bm{r})}-\frac{\delta\sum_j\mu_j h_j}{\delta W^*(\bm{r})}=0,
\end{equation}
with the orthonormal constrains
\begin{equation}
h_j=\int{d\bm{r}W^*(\bm{r})W(\bm{r}-\bm{r}_j)}=\delta_{0,j}\,.
\end{equation}
$\mu_j$'s are the usual Lagrangian multipliers. 
With straightforward computation, a nonlinear equation was obtained for the interacting Wannier functions~\cite{BW CPL} 

\begin{equation}
\begin{aligned}
&\sum_j\mu_jW(\bm{r}-\bm{r}_j)=\sum_{j_1,j_2}\langle\hat{a}_{j_1}^\dagger\hat{a}_{j_2}\rangle H_0W(\bm{r}+\bm{r}_{j_1}-\bm{r}_{j_2})\\
&+g_0\sum_{j_1j_2}^{j_3j_4}\langle\hat{a}_{j_1}^\dagger\hat{a}_{j_2}^\dagger\hat{a}_{j_3}\hat{a}_{j_4}\rangle W^*(\bm{r}+\bm{r}_{j_2}-\bm{r}_{j_1}) \\
&\times W(\bm{r}+\bm{r}_{j_2}-\bm{r}_{j_4})W(\bm{r}+\bm{r}_{j_2}-\bm{r}_{j_3}).\\
\label{wannier}
\end{aligned}
\end{equation}
where $\langle \cdot\rangle$ represents averaging over the ground state of the system. 
The ground state is to be found with the Bose-Hubbard model
\begin{equation}
\hat{H}_h=-J\sum_{\langle ij\rangle}\hat{a}_i^\dagger \hat{a}_j+\frac{U}{2}\sum_i 
\hat{a}_i^\dagger \hat{a}_i(\hat{a}_i^\dagger \hat{a}_i-1)\,,
\label{hubbard}
\end{equation}
where 
\begin{equation}
J=-\int d\bm{r}W^{*}(\bm{r}-\bm{r}_{j})H_{0}W(\bm{r}-\bm{r}_{j-1})\,
\end{equation}
and 
\begin{equation}\label{eq}
U=g_0\int d\bm{r}|W(\bm{r})|^{4}\,.
\end{equation}
Eq.(\ref{wannier}) and Eq.(\ref{hubbard}) need to be solved self-consistently together
to find the best Wannier function $W(\bm{r})$. 

\begin{figure}
  \includegraphics[height=5.5cm]{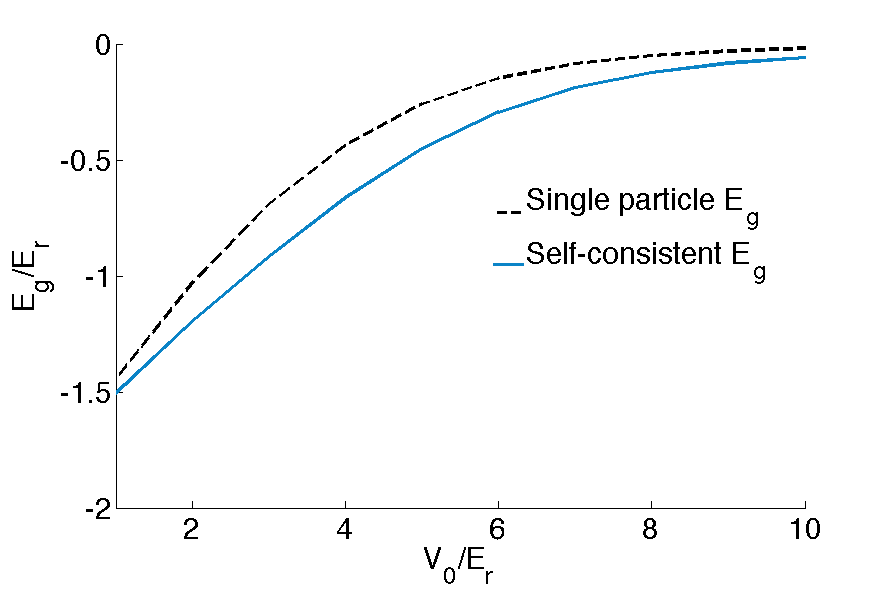}
  \caption{Comparison of the ground state energies of Bose-Hubbard model between the usual single-particle method and our self-consistent method.  The self-consistent method produces a lower ground state energy. $E_r=\hbar^2k_L^2/2m$ is the recoil energy. $k_L$ is the wave vector of optical lattice laser.}
  \label{energy}
\end{figure}

\begin{figure}[htbp]
  \centering
  \includegraphics[height=5.5cm]{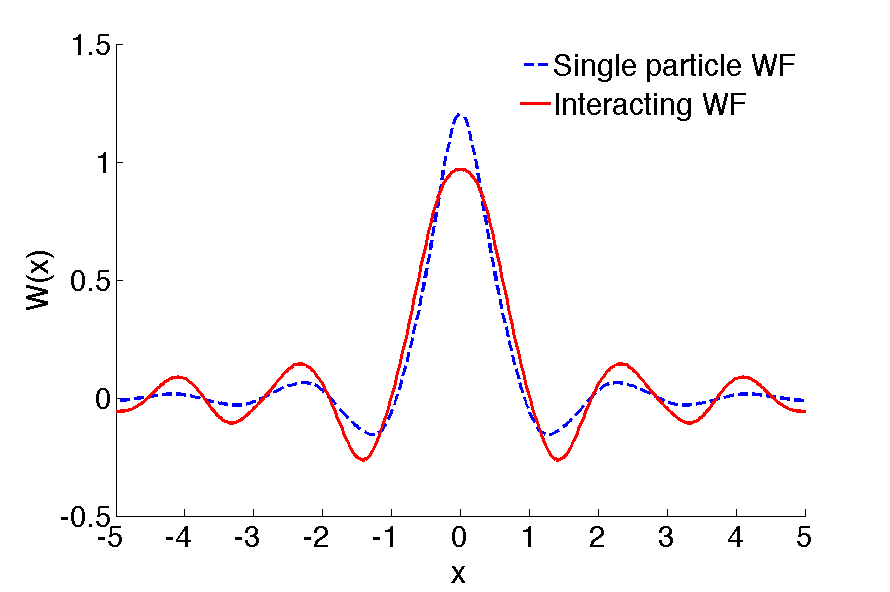}
  \caption{Blue dashed line is single particle Wannier function, and red line is interacting Wannier function by self-consistent method. $x$ is in unit of $\lambda/2$, where $\lambda$ is the wave length of optical lattice laser.}
  \label{WannierF}
\end{figure}

As an example, we consider a one dimensional six-site Bose-Hubbard model with one atom per site so that we can find its ground state with exact diagnalization\cite{JZ EJP}. 
We use two different ways to compute $J$ and $U$ in the model: (1) with the single particle
Wannier function; (2) with the interacting Wannier function obtained self-consistently with 
Eqs.(\ref{wannier},\ref{hubbard}). The ground state energy of this Bose-Hubbard model
is compared for these two methods in Fig. \ref{energy}.  The energy  computed with
the self-consistent method is indeed lower. 
Fig. \ref{WannierF} shows one  Wannier function that we obtained in superfluid regime, which is apparently broadened due to the interaction. These broadened Wannier functions can influence the tunneling parameter $J$ and on-site interaction $U$ in the single band Bose-Hubbard model.
In the following two sections, we shall compute the broadened Wannier function in
both superfluid regime and Mott regime for one dimensional systems.

\begin{figure*}
  \includegraphics[height=10cm]{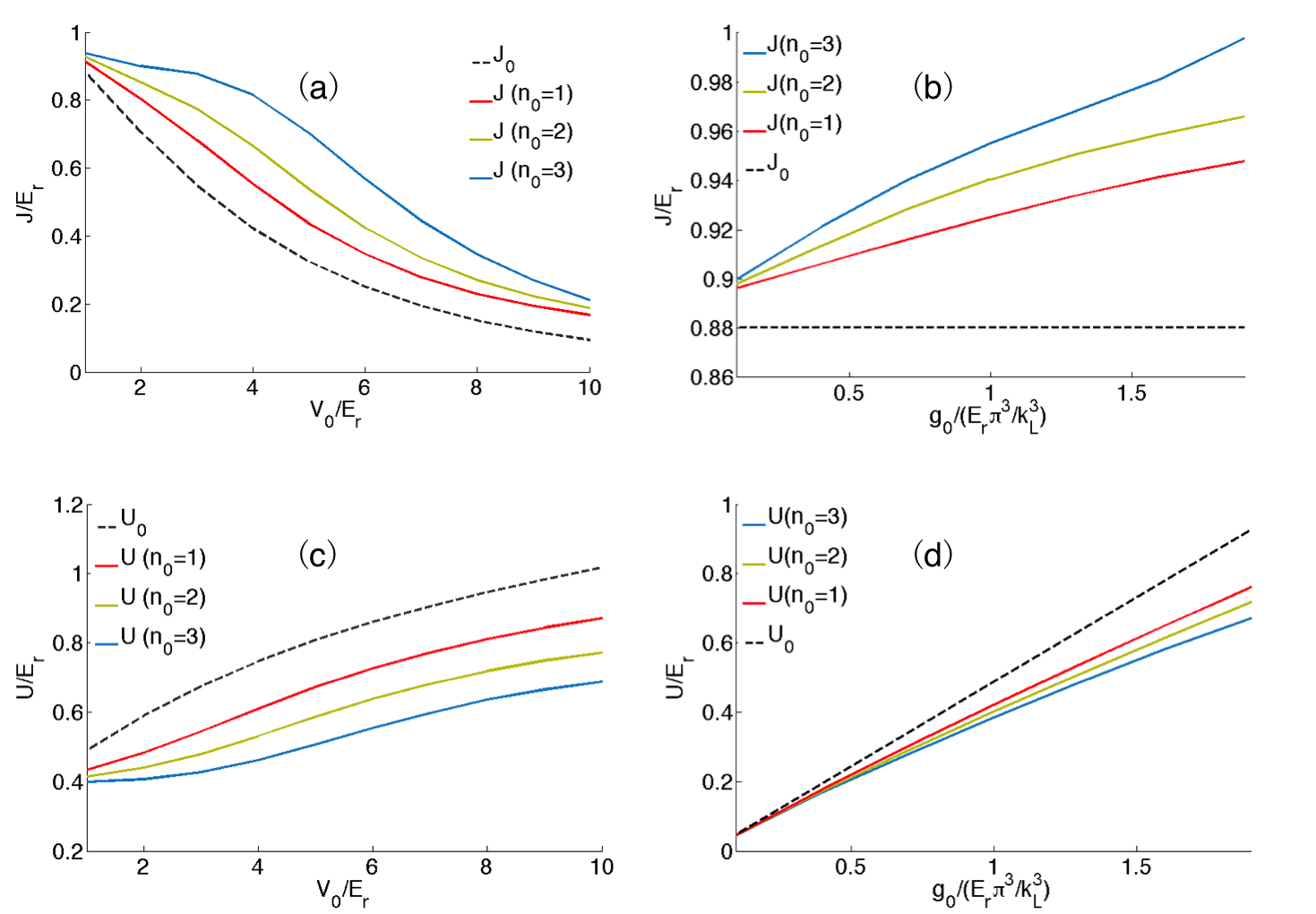}
  \caption{Superfluid regime. (a) Change of tunneling parameter $J$  with potential depth 
  for different $n_0$; (b) Change of $J$  with interaction strength for different $n_0$; (c) Change of on-site interaction $U$ with potential depth for different $n_0$; (d) 
Change of $U$ with interaction strength for different $n_0$.}
  \label{BlochBasis}
\end{figure*}

\section{Superfluid regime}

In the superfluid regime all the particles are condensed into the ground state of the system which is
a Bloch state; it is more convenient 
to use the Bloch basis. The nonlinear equations for Bloch waves are~\cite{BW CPL}
\begin{equation}
\begin{aligned}
&\tilde{\nu}_{\bm{k}}\psi_{\bm{k}}(\bm{r}) = H_0\psi_{\bm{k}}(\bm{r})\\
&+g_0\sum_{\langle{{\bm{k}}_1{\bm{k}}{\bm{k}}_3{\bm{k}}_4}\rangle}P_{{\bm{k}}_1{\bm{k}}{\bm{k}}_3{\bm{k}}_4}
\psi_{{\bm{k}}_1}^*(\bm{r})\psi_{{\bm{k}}_3}(\bm{r})\psi_{{\bm{k}}_4}(\bm{r})\,,\\
\label{bloch}
\end{aligned}
\end{equation}
where $P_{{\bm{k}}_1{\bm{k}}{\bm{k}}_3{\bm{k}}_4}=\langle\hat{b}_{{\bm{k}}_1}^\dagger\hat{b}_{\bm{k}}^\dagger\hat{b}_{{\bm{k}}_3}\hat{b}_{{\bm{k}}_4}\rangle/\langle\hat{b}_{\bm{k}}^\dagger\hat{b}_{\bm{k}}\rangle$ with $\hat{b}_{\bm{k}}=\frac{1}{\sqrt{N}}\sum_j\hat{a}_ne^{-i{\bm{k}}\cdot\bm{r}_j}$. 
In the superfluid phase, the Bogoliubov mean-field theory\cite{CP CAM} can be used to determine
and  compute $P_{{\bm{k}}_1{\bm{k}}{\bm{k}}_3{\bm{k}}_4}$ and other 
coefficients in Eq.(\ref{bloch})~\cite{BW CPL}. There is no indeed to solve Eq.(\ref{hubbard}).

In the computation, we expand the Bloch function $\psi_{\bm{k}}$ with plane waves,
\begin{equation}
\psi_{\bm{k}}(\bm{r})=\frac{1}{\sqrt{N\Omega}}\sum_{\bm{K}}a({\bm{k}}+{\bm{K}})e^{i({\bm{k}}+{\bm{K}})\bm{r}}\,,
\end{equation}
where $\Omega$ is the volume of a cell and $N$ is the number of cells. 
After plugging the above equation into Eq.(\ref{bloch}), we will obtain a set of 
nonlinear equations for $a({\bm{k}})$. We solve these nonlinear equations 
numerically and then  construct the Wannier functions of different energy bands 
by carefully choosing the phases of all the Bloch functions with Kohn's method\cite{WK PR}.

Fig. \ref{BlochBasis} shows that $J$ and $U$ change with lattice strength $V_0$ and interaction parameter $g_0$ for 
different mean particle number on a single site $n_0$. 
It is apparent that the atomic number per site
does not change the overall trends but does change $J$ and $U$ significantly. 
As the result of broadening of Wannier functions, $J$ increases with $n_0$ while $U$ decreases. 
In the range that we show in Fig. \ref{BlochBasis}, $J$ with our approach can change up to $100\%$ compared to that of a single particle result at $n_0=3$ 
while $U$ can change up to $30\%$.

\section{Mott Insulator regime}
In deep Mott-insulator regime the ground state can be approximated with $|n_0,n_0,...,n_0\rangle$, where $n_0$ is atom number per cite. As a result, the nonlinear equation for interacting Wannier function in the Mott regime is simplified to

\begin{equation}
\begin{aligned}
\sum_j\frac{\mu_j}{N_0}W(\bm{r}-\bm{r}_j) &= H_0W(\bm{r}) + g_0(n_0-1)|W(\bm{r})|^2W(\bm{r})\\
&+ 2g_0n_0\sum_{\bm{r}_j\neq0}|W(\bm{r}-\bm{r}_j)|^2W(\bm{r}).
\end{aligned}
\label{Mott}
\end{equation}

\begin{figure*}
  \includegraphics[height=10cm]{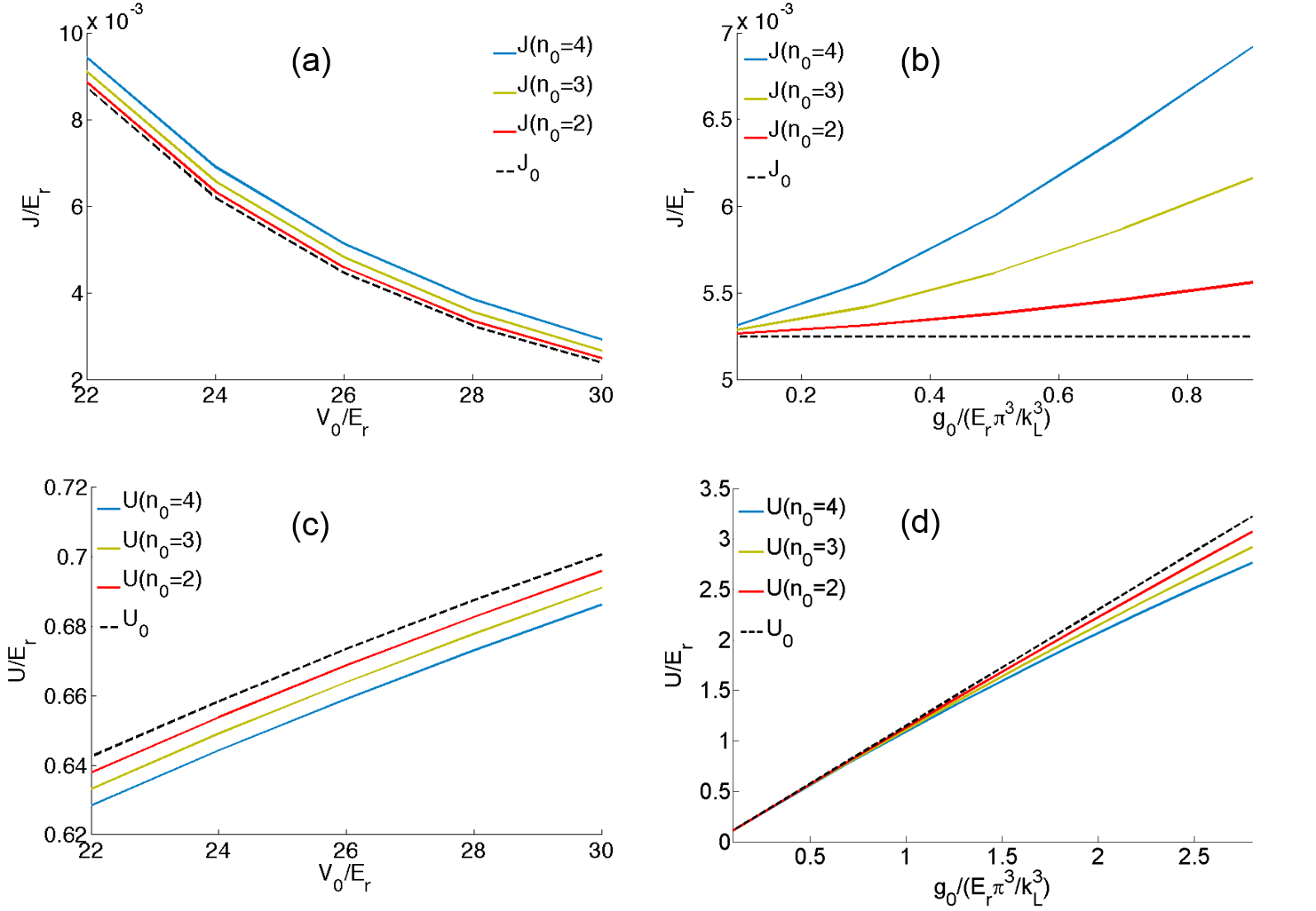}
  \caption{Mott insulator regime. (a) Change of tunneling parameter $J$  with potential depth 
  for different $n_0$; (b) Change of $J$  with interaction strength for different $n_0$; (c) Change of on-site interaction $U$ with potential depth for different $n_0$; (d) 
Change of $U$ with interaction strength for different $n_0$.}
  \label{WannierBasis}
\end{figure*}

We expand the Wannier function in terms of single-particle  Wannier functions on the same site and its nearest neighbors,

\begin{equation}
\begin{aligned}
W(\bm{r}-\bm{r}_{j})=&\sum_{n=1}^{M}[c_{n}w_{n}(\bm{r}-\bm{r}_{j-1})\\
&+b_{n}w_{n}(\bm{r}-\bm{r}_{j})+c_{n}w_{n}(\bm{r}-\bm{r}_{j+1})],\\
\end{aligned}
\end{equation}
where $w_n(\bm{r}-\bm{r}_j)$ is the single particle Wannier function for  band $n$ and  
site $j$. In this numerical calculation, we set $M=3$ and dimension for this section is 1D. 
With this expansion, we solve Eq.(\ref{Mott}) numerically to find the
interaction-broadened  Wannier functions, and compute $J$ and $U$. The results are shown 
in Fig. \ref{WannierBasis}.  It is clear that the general trends that  $J$ and $U$ 
change with $g_0$ and $V_0$ in the Mott regime are similar to the ones 
in the superfluid regime. However, there are differences. Specifically, 
as shown in Fig. \ref{WannierBasis}(a,c), the change of both $J$ and $U$ with $n_0$
has little dependence on the lattice depth $V_0$. 
As shown in Fig. \ref{WannierBasis}, $J$ can change up to $32\%$ in our self-consistent approach at $n_0=4$ compared to that of single particle Wannier function 
while $U$ can change up to $14\%$.

\section{Explanation of non-uniform clock shift}
\begin{figure}[tbp]
  \centering
  \includegraphics[height=5.5cm]{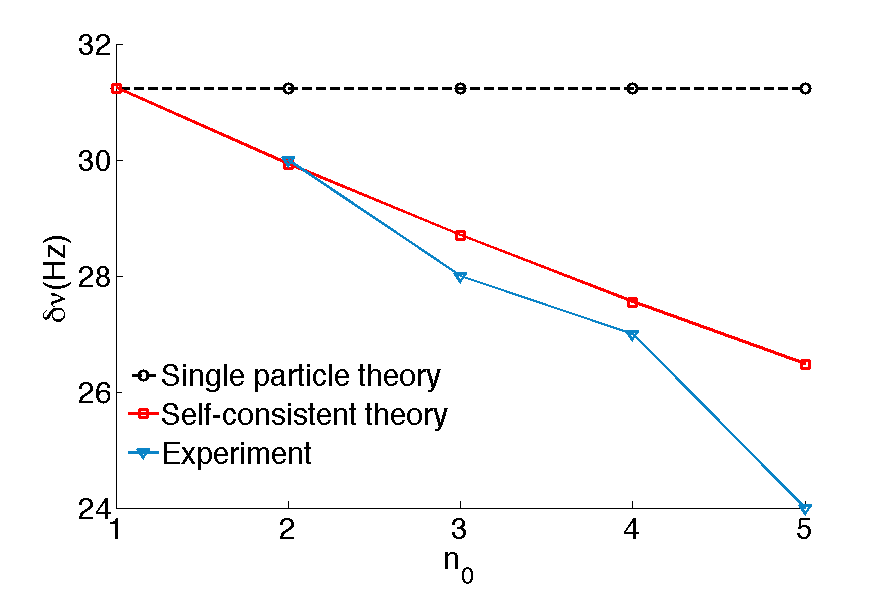}
  \caption{Clock shifts for different  atomic occupation numbers  per site. The shift becomes smaller for higher  occupation.}
  \label{shift}
\end{figure}

In atomic clock measurement, the clock frequency can shift due to collision of atoms. 
In an  experiment reported in Ref.~\cite{GC Science},  the atomic clock shift of $^{87}Rb$ was measured. In the experiment,  a $^{87}Rb$ Bose-Einstein condensate was prepared in the 
$|F=1,m_f=-1\rangle$ state and loaded into a 3D optical lattice. With the  increase of 
the lattice depth, the system changed from superfluid phase to Mott insulator (MI) phase. 
Due to the trapping potential,  the atomic gases was separated into MI shells each of which
has a different occupation number $n_0$ per site. One can use radio wave to excite atoms in $F=1$ state  to $F=2$ state. In different hyperfine states, the scattering lengths between atoms are different. Therefore, the atoms transferred to the $F=2$ state have a slightly different mean field energy; this can cause a clock frequency shift, which  is given by~\cite{CP CAM, DH PRA}
\begin{equation}
\delta \nu=\frac{U}{h}(a_{21}-a_{11})/a_{11}\,, 
\end{equation}
where $a_{11}$ and $a_{22}$ are scattering lengths for atoms in $F=1$ 
and $F=2$ state, respectively, and $a_{12}$ is scattering length between atom in $F=1$ state and atom in $F=2$ state. If the on-site interaction $U$ is calculated with single-particle Wannier 
functions, this clock shift is independent of $n_0$, the number of atoms per site. 
However, it was observed in the experiment, the clock shift decreases with $n_0$ as 
shown in Fig. \ref{shift}.  

In our view, this decrease of clock frequency shift in Fig. \ref{shift} is 
caused by the broadening of Wannier function: when $n_0$ increases, 
the Wannier function becomes broader and $U$ decreases as shown in Fig. \ref{WannierBasis}(c,d);
consequently, the clock shift decreases. With our self-consistent method, we calculated the 3D interacting Wannier function for different atom number $n_0$ per site in 3D Mott-insulator regime,
and eventually $U$ and the clock shift $\delta\nu$. The results are compared to 
the experimental results in Fig. \ref{shift}; there is a very good agreement.

\section{Conclusion}

With a self-consistent theory developed earlier, we have computed 
the effect of interaction on single-band Wannier function. 
In particular, we have considered both superfluid regime and Mott regime. We  found
that as the result of broadening of Wannier function by interaction, 
the tunneling parameter $J$ and $U$ can change significantly. Our theory 
was applied to a clock shift experiment: a very good agreement was found
between our theoretical results and experimental results. 
The regime near  the transition from superfluid to Mott insulator is 
not studied in this work and will be investigated in the future. 

\section{Acknowledgement}
We thank Qizhong Zhu for helpful discussions. This work is supported by the NBRP of China (2013CB921903,2012CB921300) and the NSF of China (11274024,11334001,11429402).


\begin{thebibliography}{}
\bibitem[1]{GW PR} G. H. Wannier, Phys. Rev. {\bf 52} 191 (1937).
\bibitem[2]{NA SSP} N. W. Ashcroft and N.D. Mermin \emph{Solid State Phys. Sanders, Philadelphia}(1976).
\bibitem[3]{NM PRB} N. Maezari and D.Vanderbilt Phys. Rev. B {\bf 56},12847 (1997).
\bibitem[4]{WK PRB} W. Kohn, Phys. Rev. B {\bf 7}, 4388 (1973).
\bibitem[5]{IB Nature} M. Greiner, O. Mandel, T. W. Hansch, I. Bloch, Nature {\bf 415}, 39 (2002).
\bibitem[6]{IB RMP} I. Bloch, J. Dalibard, W. Zwerger, Rev. Mod. Phys. {\bf 80}, 885 (2008).
\bibitem[7]{DJ PRL} D. Jaksch, C. Bruder, J. I. Cirac, C. W. Gardiner, P. Zoller, Phys. Rev. Lett. {\bf 81}, 3108 (1998).
\bibitem[8]{JL NJP} J. B. Li, Y. Yu, A. M. Dudarev, Q. Niu, New J. Phys. {\bf 84}, 154 (2006).
\bibitem[9]{ZL arXiv} Z. X. Liang, B. B. Hu, B. Wu, arXiv:0903.4058 (2009).
\bibitem[10]{PS PRA} P. I. Schneider, S. Grishkevich, A. Saenz, Phys. Rev. A {\bf 80}, 013404 (2009).
\bibitem[11]{DM PRA} D. Masiello, S. B. McKagan, W. P. Reinhardt, Phys. Rev. A {\bf 72}, 063624 (2005).
\bibitem[12]{VY LP} V. Yukalov, Laser Phys. {\bf 19}, 1 (2009).
\bibitem[13]{DL NJP} D. S. Luhmann, O. Jurgensen, K. Sengstock, New J. Phys. {\bf 14}, 033021 (2012).
\bibitem[14]{KG PRL} K. Gibble, S. Chu, Phys. Rev. Lett. {\bf 70}, 1771 (1993).
\bibitem[15]{GC Science} G. K. Campbell \emph{et al}, Science {\bf 313}, 649 (2006).
\bibitem[16]{GC2 Science} G. K. Campbell \emph{et al}, Science {\bf 324}, 360 (2009).
\bibitem[17]{MM Science} M. J. Martin, \emph{et al}, Science {\bf 341}, 632 (2013).
\bibitem[18]{BW CPL} B. Wu, Y. Xu, L. Dong, J. R. Shi, Chin. Phys. Lett. {\bf 29}, 083701 (2012).
\bibitem[19]{AS PRA} A. I. Streltsov, O. E. Alon, L. S. Cederbaum, Phys. Rev. A {\bf 73}, 063626 (2006).
\bibitem[20]{OA PRA} O. E. Alon, A. I. Streltsov, L. S. Cederbaum, Phys. Rev. A {\bf 77}, 033613 (2008).
\bibitem[21]{KS NJP} K. Sakmann, A. I. Streltsov, O. E Alon, L. S. Cederbaum1, New J. Phys. {\bf 13}, 043003 (2011).
\bibitem[22]{JZ EJP} J. M. Zhang, R. X. Dong, Eur. J. Phys. {\bf 31}, 591 (2010).
\bibitem[23]{CP CAM} C. J. Pethick, H. Smith, \emph{Bose-Einstein Condensation in Dilute Gases, Cambridge}, 2nd Edition (2008).
\bibitem[24]{WK PR} W. Kohn, Phys. Rev. {\bf 115}, 809 (1959).
\bibitem[25]{DH PRA} D. M. Harber, H. J. Lewandowski, J. M. McGuirk, E. A. Cornell, Phys. Rev. A. {\bf 66}, 053616 (2002).
\bibitem[26]{GB PRL} G. Baym, C. J. Pethick, Phys. Rev. Lett. {\bf 76}, 6 (1996).
\end{thebibliography}
\end{document}